\documentclass[superscriptaddress,amsmath,amssymb,aps,prb,twocolumn]{revtex4-2}

\usepackage{comment}
\usepackage{graphicx}
\usepackage{dcolumn}
\usepackage{bm}
\usepackage{amssymb}
\usepackage{amsmath}
\usepackage{physics}
\usepackage{sublabel}

\usepackage[utf8]{inputenc}
\usepackage{hyperref}
\usepackage[usenames, dvipsnames]{color}
\usepackage[english]{babel}
\usepackage[T1]{fontenc}
\usepackage{bbm}

\usepackage{float}
\usepackage{verbatim}
\hypersetup{colorlinks = true, linkcolor=blue, citecolor=blue, urlcolor=blue}
\usepackage{xcolor}

\begin{document}

\preprint{APS/123-QED}
\title{Inclined junctions in monolayer graphene: A gateway toward tailoring valley polarization of Dirac fermions}
%\title{Modeling of valley splitter using homo and hetero tilted junctions in monolayer graphene}
%\title{Tunable multi-terminal graphene nanoribbon valley-selective devices}
\author{Shrushti Tapar}
\author{Bhaskaran Muralidharan}%
 \email{bm@ee.iitb.ac.in}
\affiliation{Department of Electrical Engineering, Indian Institute of Technology Bombay, Powai, Mumbai-400076, India}
%\date{\today}% It is always \today, today, %  but any date may be explicitly specified
\begin{abstract}
%Generating discernible valley contrasts and segregating valley-indexed fermions in real space within graphene poses considerable challenges due to its inherent lattice inversion symmetry. This study unveils an interesting finding: introducing valley contrast through anisotropic chiral transport in isotropic Dirac systems like graphene, achieved by implementing a tilted PN junction (PNJ). Remarkably, this method maintains the exceptional mobility of fermions and the density of states, offering distinct advantages over alternative strategies. The exploration into optimizing the experimental setup for desired device performance delves into intricacies. Evaluating the sequence of the doped region emerging from the source demonstrates its substantial influence on valley-index fermion refraction across the potential interface. Additionally, critical parameter variations such as the tilt and transition width across the junction have been examined. The variations in transition width result in increased transmission, which is counterintuitive and attributed to speculative edge scattering. Moreover, this study delves into the impact of Anderson's short-range edge disorder on device performance. The observed chiral anisotropic transport mirrors behaviors found in Dirac-Weyl semimetals, linked to their characteristic tilted Dirac band structure. Consequently, by implementing a tilted PNJ, non-tilted Dirac systems can be artificially tuned into tilted counterparts, enabling the induction of anisotropic transport.\\
Generating discernible valley contrasts and segregating valley-indexed fermions in real space within graphene poses considerable challenges due to the isotropic transport within the continuum energy range for degenerate valleys. This study unveils an interesting finding: introducing valley contrast through anisotropic chiral transport in isotropic Dirac systems like graphene, achieved by implementing a tilted PN junction. The tilted junction shifts the angular spectrum to larger angles in accordance with the tilt angle. This modifies the pseudospin-conserved modes across the junction, resulting in valley-resolved chiral transport. This approach not only induces valley splitting within the real space but also preserves the remarkable mobility of fermions, offering distinct advantages over alternative strategies. The comprehensive analysis includes optimizing the experimental setup, scrutinizing factors such as the sequence of the doped region, and examining critical parameters like the tilt angle ($\delta$) and transition width ($d$) across the junction. Surprisingly, an increased transition width enhances transmission, attributed to specular edge scattering. Importantly, the system remains resilient to Anderson’s short-range edge disorder. The broader implication lies in the transformative potential of inducing analogous anisotropic chiral transport behaviors in isotropic Dirac systems, resembling the characteristics of tilted Dirac-Weyl semimetals, by incorporating a tilted PNJ.
\end{abstract}


%\keywords{Suggested keywords}%Use the 22show keys class option if the %display desired
\maketitle
 
%\tableofcontents

\section{\label{sec:level1}Introduction}
\indent Continuous technological advancements have opened avenues to exploit device functionalities, from fundamental physical phenomena like spin \cite{HIROHATA2020166711, lin2019two,JOSHI20161503}, topology \cite{groning2018engineering, tokura2019magnetic, jana2022robust, Jana_PRApp}, and valleys \cite{schaibley2016valleytronics,vitale2018valleytronics,ang2017valleytronics,schirber2021valley,qiao2014quantum} to intricate quantum states \cite{makhlin2001quantum, nakamura1999coherent, pusey2012reality} and emerging concepts like skyrmions \cite{back20202020, hamamoto2015quantized}. Recently, there has been a focused effort to investigate and take use of the valley degree of freedom \cite{vitale2018valleytronics,ang2017valleytronics}, which in some materials produces encodable and permanent valley states. Encoding valley states requires discernible differences in charge occupancy between distinct valleys, termed `valley contrast'. Valley polarization, fundamental in valleytronics, relies on this measurable valley contrast to enable encoding and manipulation of valley states.\\
\indent Achieving valley contrast in graphene faces challenges due to the simultaneous presence of inversion and time-reversal symmetries. Despite this, graphene remains a well-studied system. Staggered sublattice potentials \cite{zhou2007substrate,PhysRevLett.127.116402} helps in breaking the inversion symmetry \cite{xiao2007valley,qiao2011spin,asmar2017minimal} and lifting valley degeneracy. Higher-energy bands in graphene exhibit trigonal Fermi surfaces oriented differently for distinct valleys, enabling anisotropic valley-specific transport \cite{pereira2008valley,garcia2008fully}. Line defects in monolayer graphene act as semi-transparent scattering centers, creating a valley filter based on carrier valley-index \cite{liu2013controllable,gunlycke2011graphene}. Although uniaxial strain helps separate electron trajectories, it cannot cause valley-polarized conductance \cite{chauwin2022strain,wu2011valley}. In order to achieve this, transmission symmetry must be broken by a magnetic barrier, allowing valley-polarized transport \cite{yesilyurt2016perfect,zhai2010magnetic,zhai2012valley,lu2016valley,milovanovic2016strain,cavalcante2016all}. A non-uniform strain could generate local valley contrast through a pseudomagnetic field \cite{zhai2018local,wu2018quantum,low2010strain}.However, the it does not efficiently convert locally achieved valley contrast into valley-resolved conductance \cite{hsu2020nanoscale,li2020valley,settnes2017valley}.\\
\indent Establishing valley contrast through electrostatic barriers is pivotal for on-chip functionalities \cite{wang2017valley}. Earlier investigations have detailed the phenomenon of chirality-induced anisotropic refraction occurring at the interfaces of electrostatic barriers in Weyl semimetals \cite{yesilyurt2019electrically, PhysRevB.97.235113, zhou2019valley}. This is attributed to the tilted Dirac cones in Weyl semimetals. Our work introduces a similar concept tailored for monolayer graphene, which possesses a non-tilted Dirac cone. The proposed adaptation involves incorporating a tilted PN junction designed using a split-gate geometry \cite{PhysRevB.80.155406,yang2012conductance,sutar2012angle, sajjad2012manifestation}.\\
\indent Given the relativistic nature, fermions in graphene exhibit optical-like properties following Snell's-Descartes law across the potential barrier interface \cite{lee2015observation,darancet2009coherent,reijnders2017diffraction,cheianov2007focusing}. The intriguing transport property of Klein tunneling is evident due to the pseudo-spin conservation across the junction \cite{allain2011klein, katsnelson2006chiral, beenakker2008colloquium}. When a tilted PN junction is introduced in graphene, the degenerate valley modes break into different pseudo-spin-conserved modes when they refract across the potential interface \cite{PhysRevB.80.155406}. This phenomenon leads to valley-specific refraction across the potential barrier.\\
\indent In this study, we explore the mechanism behind valley separation in monolayer graphene using a tilted PN junction. Through the tilted PN junction, the proposed study reveals the possibility of introducing anisotropic chiral Klein tunneling in Dirac isotropic systems. The extension of the tilted PN junction concept to Dirac isotropic systems not only introduces anisotropy but also amplifies anisotropic behavior within tilted Dirac systems. This method promises efficient valley splitting and straightforward implementation. Our objective is to ascertain the ideal circumstances that produce the intended valley splitting effect, all within the bounds of experimental feasibility. This entails examining the impact of various parameters on valley separation across the PN junction, including the tilt angle, the transition width, and the sequence of the doped region that resembles different refractive indices in an optical system. Additionally, we examine the impact of Anderson’s edge disorders on valley polarization.\\

\begin{figure}[!htbp]
	\includegraphics[width=\linewidth]{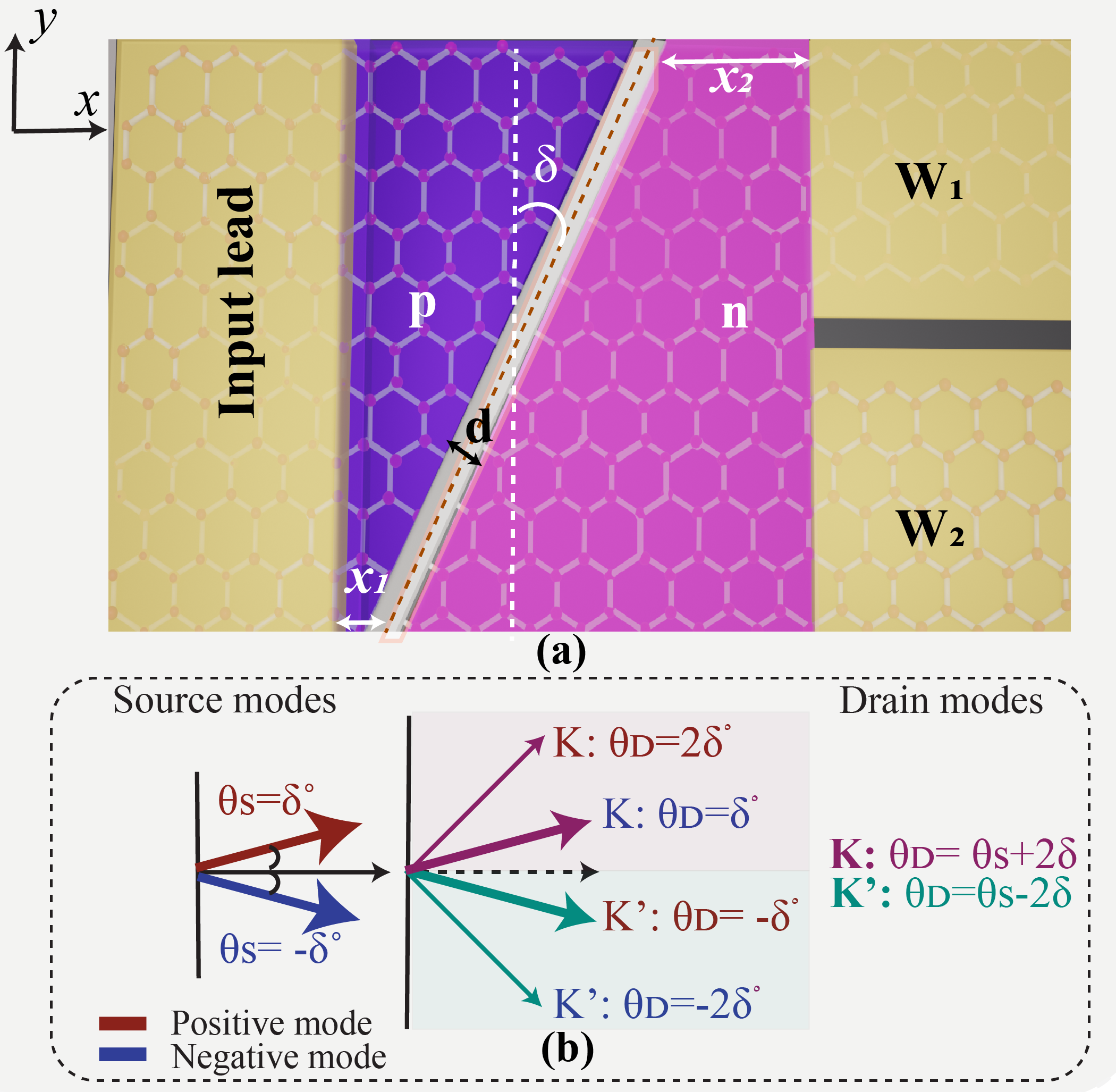}
	\caption{Schematic and the working principle: (a) The proposed multi-terminal graphene device consists of an input lead and two output leads, $W_1$ and $W_2$. The scattering region is electrostatically doped using split gates for inducing the tilted junction. This junction, formed between the $p$ and $n$ doped regions, is angled at $\delta$ with respect to the $y-axis$. The potential undergoes a gradual change across the doped regions, smoothly varying over the transition width, $d$. (b) Pseudo conserve modes for a tilted junction: When the junction is tilted with an angle $\delta$, following the Klein tunneling the maximum contributing source modes $\theta_S = 0^{\circ}$ shifts to $\theta_S =\pm \delta^{\circ}$ modes shown in red and blue colored arrows respectively. When the incoming modes incident on a tilted junction they get scattered at different angles depending upon their valley index $K$ and $K'$. The scattered source modes get transmitted to drain modes followed by the relations shown. The maximum contributing source modes $\pm\delta^{\circ}$ split angular spectrum on the drain side in the domains dominated by $K$ and $K'$valley components.}
    \label{fig:Fig_1}
\end{figure} 
\indent The paper is structured as follows: Section \ref{Sec2} presents a comprehensive overview of the device modeling and simulation. In Section \ref{Sec3}, the key results are discussed. Section \ref{Sec3}A covers the analysis of Dirac fermion transport across both bipolar and unipolar junctions, while Section \ref{Sec3}B delves into the discussion of chiral tunneling across normal and tilted junctions. This section intricately examines the operational principles governing valley separation in the tilted PNJ, building upon the established framework of normal PNJ transport. Furthermore, Section \ref{Sec3}C highlights the impact of varying transition width and tilt angle on the degree of valley polarization. To assess the impact of edge disorder, we introduced a short-range Anderson disorder and analyzed its influence on valley polarization in \ref{Sec3}D. Finally, Section \ref{Sec4} encapsulates the main conclusions drawn from the study and outlines future prospects.

\section{Device and simulations details} \label{Sec2}
\indent The proposed multi-terminal graphene device features one input and two output leads, denoted as $W_1$ and $W_2$ as depicted in Fig.\ref{fig:Fig_1}(a) For the whole simulation, the input/output leads of the device and scattering region are modeled using an anti-zigzag edge graphene configuration. An anti-zigzag ribbon is characterized by the presence of an odd number of zigzag atomic rows arranged within the graphene structure. The scattering region is electrostatically doped to form a tilted junction using split gate geometry. This junction, occurring between the doped $p$ and $n$ regions, is inclined at an angle $\delta$ with respect to the $y-axis$. The potential gradually changes across the doped areas, displaying a smooth transition over the width, $d$. \\
\indent In terms of dimensions, the input lead and scattering region have a width of $W = 100~nm$, whereas the widths of the two output leads, $W_1$ and $W_2$, are approximately half that of the input lead. Depending on the tilt angle $\delta$, the length of the scattering region ranges from $200~nm$ to $300~nm$. The lower end of a junction is situated at a distance of $x_1 = 10-15~nm$ from the input side, marking the initiation point of the potential transition. Meanwhile, the opposing corner of the junction is positioned minimum at $x_2 = 50-70~nm$ from the drain end.\\ 
\indent The symmetric PNJ is formed by an equal level of electrostatic doping on both sides of the junction. The Fermi level lies $0.3~eV$ below the Dirac point for the $p$ region and above for the $n$ region. The polarization is adjustable using various parameters such as the transition width $d$, tilt angle $\delta$, and the position of junction from both the source and drain sides. It is noteworthy that the anti-zigzag graphene arrangement of the proposed device naturally produces the valley filter effect at Fermi crossing energy \cite{akhmerov2008theory,rycerz2007valley}.\\
\indent The simulations utilize the scattering matrix formalism \cite{groth2014kwant}. This approach is integrated with the Landauer-Büttiker formalism, where the structure is partitioned into discrete sections along its length, ensuring ballistic transport across consecutive segments. Each section forms a scattering matrix that links incoming wave amplitudes to outgoing waves for individual modes. The collective merging of these matrices generates the overall scattering matrix for the entire structure. From this matrix, transmission amplitudes for all propagating modes are directly derived. These amplitudes then facilitate the computation of current under both linear and non-linear response regimes, presuming phase-coherent transport, as outlined in \cite{bandyopadhyay1991generalized}. The simulations where carried out using kwant\cite{groth2014kwant}. The Hamiltonian represented in the second quantization form is constructed using the tight-binding model for graphene given by, 
\begin{equation}
\hat{H} = -t \sum_{i,j} \hat{c}_i^ {\dagger} \hat{c}_j + h.c. ,
\end{equation}
where, $t$ is a hopping integral across the nearest neighbour $\braket{i,j}$ with the energy equal to $2.7~eV$.  $c^{(\dagger)}_{i}$ is an annihilation (creation) operator of an electron on-site $i$ and the value of a lattice constant $a$ is equal to $0.246~nm$. The valley resolved transmission is calculated for both the output leads $W_1$ and $W_2$, which are annotated as $T_{K_{1}}$ and $T_{K^{\prime}_{1}}$ for $K$ and $K^{\prime}$ in $W1_1$ and $T_{K_{2}}$ and $T_{K^{\prime}_{2}}$ in $W_2$, respectively. The polarization $P_{1,2}$ in the leads $W_1$ and $W_2$ is given by
\begin{eqnarray}
 P_{1,2} =\frac{T_{K_{1,2}}-T_{K^{\prime}_{1,2}}} {T_{K_{1,2}} + T_{K^{\prime}_{1,2}}}
\end{eqnarray}
Having discussed the device geometry and simulation methodology, we now delve into the underlying physics of anisotropic scattering and analyze the effects of various parameters.

\section{Results and Discussion} \label{Sec3}
\subsection{Dirac fermions transport across bipolar and unipolar junction}
\begin{figure}[!htbp]
	\includegraphics[width=\linewidth]{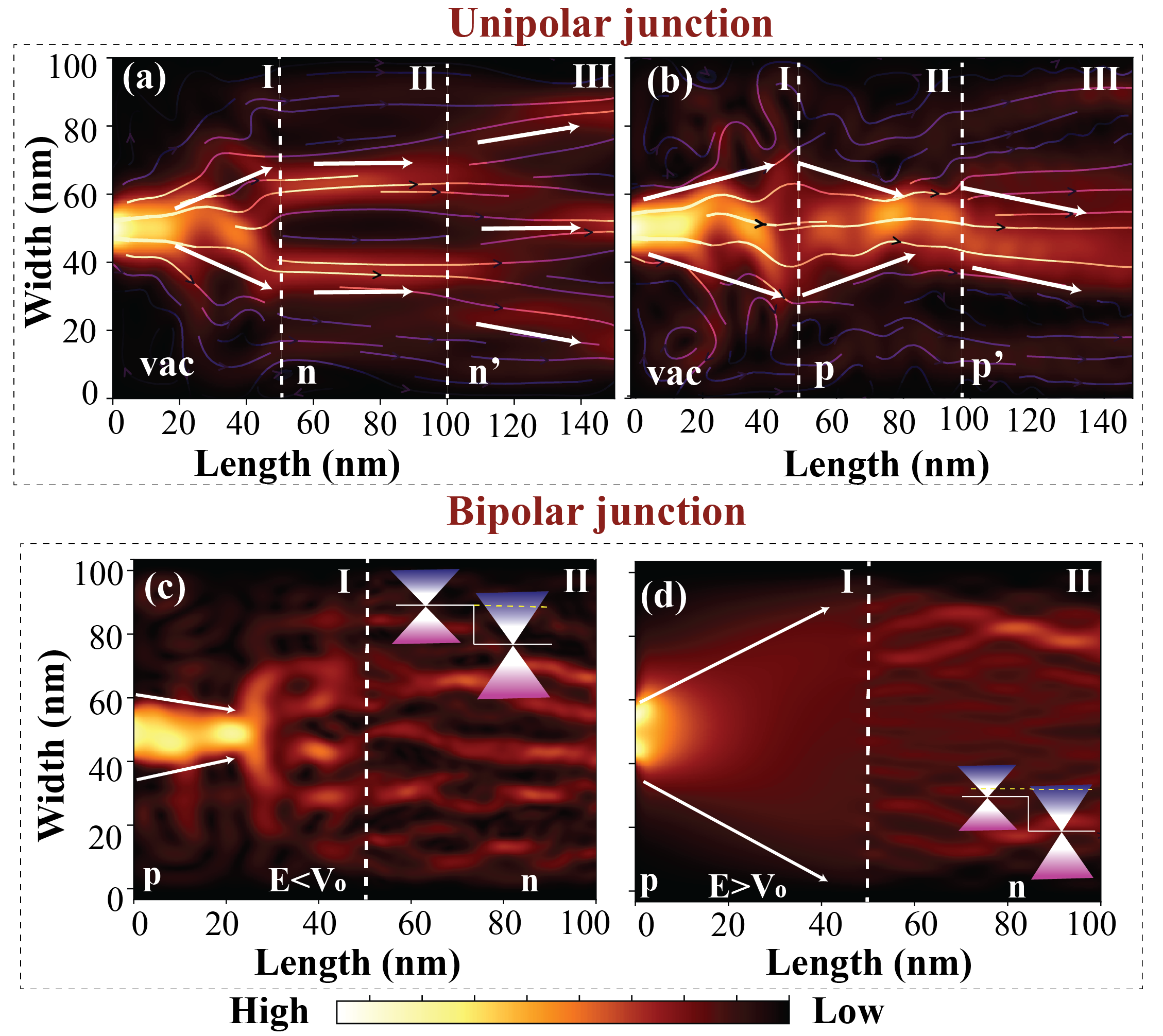}
	\caption{Current density plots for unipolar and bipolar junctions: (a) and (b) depict unipolar junctions with doped regions exhibiting escalating doping levels of the same polarity. Region I represents the vacuum region with no doping, while regions II and III demonstrate increasing doping densities of $0.2~eV$ and $0.3~eV$, respectively, marked by an abrupt junction. The distinction between (a) and (b) lies in the doping polarity, categorizing the complete device doping profile as (a) $vac-n-n^{+}$ and (b) $vac-p-p^{+}$. The incoming fermion energy is set at $0.15~eV$ in (a, b, and c), while in (d), it is $0.35~eV$. In the $n$-type UJ plot (a), the wavevectors spread out in region I, becoming further refracted towards the junction, notably prominent in region III due to the higher doping density, resulting in higher positive RI. In plot (b), the wavevectors spread in region I focus in region II due to the converging effect of negative refraction. The particle motion shifts from CB to VB from region I to region II. The prominent negative refraction in region III leads to the downward motion of the focused beam from region II, owing to the higher doping density. For BJ plots (c and d), the doping profiles and densities remain constant, differing only in the incoming fermion's energy. In plot (c), where $E< V_0$, the device functions as a BJ. In region I, the wavevectors from the input are restrained from diverging due to the presence of the $p$ region, showcasing a converging effect. As the wavevectors progress into region II, the divergence effect becomes more pronounced. Conversely, in plot (d) where $E> V_0$, the junction acts as a UJ, demonstrating CB to CB transmission. A notable disparity is observed between region I in (c) and region I in (d). In (d), the wavevectors are completely dispersed, which further diverge and exhibit interference patterns in region II.}
    \label{fig:Fig_2}
\end{figure}

\indent The lattice symmetries and band structure of graphene offer a rich playground for exotic physics. In its continuum energy range, fermions in graphene mirror relativistic particles, attributed to a linear energy-momentum dispersion and giving rise to Dirac optics. This field enables Dirac fermions to mimic optical functionalities like lenses \cite{reijnders2017diffraction, cheianov2007focusing}, microscopy \cite{boggild2017two}, splitters, reflectors \cite{graef2019corner}, mirrors \cite{sanz2020crossed}, waveguides \cite{PhysRevB.81.245431}, and meta-materials, all achievable by configuring the doping profile.\\
\indent Understanding electron transport across different doping domains is pivotal for advancing Dirac optics in graphene. Doping is achieved using split gates, creating $p-doped$ regions and $n-doped$ regions. This doping profile significantly affects the transport behavior of a device.
Depending on the polarity ($n$ and $p$) and density of doped regions, the resulting junction can be a Unipolar Junction (UJ) or a Bipolar Junction (BJ). A UJ involves intra-band transitions, while a BJ involves inter-band transitions between the valence band (VB) and the conduction band (CB) and vice versa.\\
\indent In graphene, carrier motion follows a pattern reminiscent of Snell's Law, akin to the behavior of optic rays within optical setups. Snell’s refraction law, applied in optical systems, is expressed as:
\begin{eqnarray}
n_1 sin\theta_i = n_2 sin\theta_r
\end{eqnarray}
where, $n_1$ and $n_2$ represent the refractive indices of regions on either side of the interface, dictating light speed, propagation, refraction angles, and boundary reflections.\\
\indent In an analogous graphene system, incident and refracted angles ($\theta_i$ and $\theta_r$) refer to fermion observation angles across the potential interface. In an electronic system, the refractive indices (RI) $n_i$ and $n_r$ correspond to Fermi wavevectors $k_{f1}$ and $k_{f2}$ determined by doping density and polarity. Fermion behavior at interfaces complies with an anomalous Snell’s law:
\begin{eqnarray}
\frac{sin\theta_i} {sin\theta_r} = \frac{E-V_0}{E}=\frac{n_r}{n_i}\propto \frac{k_{f2}}{k_{f1}}
\end{eqnarray}
\indent The symbol $E$ represents fermion energy, while $V_0$ denotes the potential barrier resulting from doping in graphene. The relationship is such that $n_i \propto E$, and $n_r \propto (E-V_0)$, which equates to $E_{\text{kin}}$, representing the kinetic energy. Consequently, $n \propto \alpha E_{\text{kin}}$, where $\alpha$ is the band index, positive for the conduction band (CB) and negative for the valence band (VB) \cite{allain2011klein}. Graphene exhibits characteristics of both positive and negative refractive materials, leading to ray divergence and convergence, respectively, at interfaces. In literature, researchers have leveraged the convergence effect to replicate Veselago lensing \cite{cheianov2007focusing, reijnders2017diffraction, boggild2017two}, aiming to precisely focus and manipulate carrier beams using graphene $pn$ junctions.\\
\indent Different device configurations with varied doping profiles help illustrate positive and negative refractive indices and their respective diverging and converging effects. Fig. \ref{fig:Fig_2} displays device current density plots, providing a visual explanation of these effects.\\
\indent Two junction types, Unipolar Junction (UJ) and Bipolar Junction (BJ), are analyzed. The device spans $100~nm$ in width and $150~nm$ in length, featuring a narrower input lead of $20~nm$  and output leads (not visible in the plots). In Fig. \ref{fig:Fig_2} (a) and (b), The UJ structure shows Region I, which is pristine graphene, and two doped regions that show increasing doping density and are separated by an abrupt interface. In regions II and III, doping densities are $0.2~eV$ and $0.3~eV$, respectively, with a carrier energy of $0.15~eV$. Notably, (a) and (b) differ in doping polarities, represented as $vac-n-n^{+}$ and $vac-p-p^{+}$, respectively.\\  
\indent The Current density plots for $n$-type and $p$-type UJ are depicted in Fig. \ref{fig:Fig_2}(a-b). In the case of the $n$-type UJ, a distinct diverging effect is evident in regions II and III, as seen in Fig. \ref{fig:Fig_2}(a). Here, since $E > 0$ in region I and both regions II and III are $n$-doped with the Fermi level above the fermion energy, the fermion follows a motion from the conduction band (CB) to the conduction band (CB) with lower kinetic energy (KE) and more available states in proceeding regions. This behavior mirrors positive refraction characteristics, leading to the observed diverging effect.\\
\indent The In region III of \ref{fig:Fig_2}(b), where $E_f = -0.3~eV$, the kinetic energy becomes even more negative compared to region II, resulting in a higher negative refractive index (RI) in III. Consequently, as the converged wavevectors transition from region II to region III, they undergo additional refraction in a downward direction, providing clear evidence of negative refraction.\\

\indent In region III of Fig.\ref{fig:Fig_2}(b) featuring $E_f = -0.3~eV$, the KE becomes even more negative compared to region II, leading to a higher negative RI in III. Consequently, as the converged wavevectors transition from region II to region III, they experience further refraction in a downward direction, providing clear evidence of negative refraction.\\ 
\indent The Veselago lensing effect observed in Fig.\ref{fig:Fig_2}(b) is also noticeable in a Bipolar junction featuring symmetric doping. Typically, achieving the Veselago lensing effect requires sequencing the $n$ region before the $p$ region \cite{cheianov2007focusing}. However, in this study, we are employing the reversed configuration, where the $p$ region precedes the $n$ region. This reversal results in a divergence effect on the drain side, contrary to the expected outcome. Fig.\ref{fig:Fig_2} (c-d) showcase symmetric abrupt PNJ configurations, each measuring $100~nm$ in length and width, with the junction placed precisely at the midpoint, at $50~nm$. Both the regions are equally doped at Fermi levels at $0.3~eV$.\\
\indent The distinction between Fig.\ref{fig:Fig_2} (c) and (d) pertains solely to the energy of an incoming fermion: $E=0.15 ~eV$ in (c), \textit{i.e.}, below the barrier potential and $E=~0.35~eV$ in (d) \textit{i.e.}, above the barrier potential. In Fig.\ref{fig:Fig_2} (c), within Region I, incoming wavevectors from the input lead exhibit limited dispersion. As they progress into the $p$ region, they gradually converge. Subsequently, upon transitioning to Region II, these wavevectors undergo positive refraction (divergence) as fermions move from the valence band (VB) to the conduction band (CB). There is inter-band transport from Region I to Region II.\\
\indent In Fig.\ref{fig:Fig_2}(d), with $E>E_f$ in the region I, fermions reside in the CB, reflecting positive kinetic energy, thus establishing a positive RI within the region I. This is evident from the broader dispersion of wavevectors from the input lead, noticeably wider compared to Fig.\ref{fig:Fig_2}(c). As fermions advance into region II, characterized by the $n$-doped region, CB to CB transport initiates Unipolar Junction (UJ) transport. Subsequent refraction in region II generates an interference pattern.\\
\indent The rationale behind exploring this PNJ configuration is the need for a dispersive effect on the drain side. In addition, the fermion energy in this configuration determines the BJ and UJ properties depending on how it correlates with the barrier height.\\ 
\indent The subsequent section delves into exploring chiral tunneling within a tilted junction, a pivotal factor in fostering valley polarization. Here, the substantial distinction in valley polarization can be observed based on the distinct behaviors of UJ and BJ characteristics depending on fermion energy.\\
\subsection{Chiral tunneling across normal and tilted junctions: }
\indent The transport phenomena through a doped junction in graphene exhibit a fascinating attribute known as Klein tunneling. \cite{allain2011klein,beenakker2008colloquium,katsnelson2006chiral}. This peculiar behavior entails a fermion experiencing unit transmission when incident normally, regardless of the height and width of the potential barrier. This phenomenon is rooted in the conservation of pseudo-spin across the interface, resulting in no backscattering for normally incident fermions. Moreover, in bipolar junctions (BJ), transport becomes highly selective concerning angles, suppressing angular components beyond a critical angle. This critical angle emerges when the carrier energy ($E$) equals half the potential barrier energy ($V_0$), reflecting higher angular components as $E$ increases.\\
\indent The interface within a PNJ can be structured in two ways: abrupt or smooth. In an abrupt junction, the potential undergoes significant changes within a length scale smaller than the lattice constant $a$. Conversely, in a smooth junction, the potential profile gradually varies across a length scale. This varying width is termed the transition width $d$. Within a smooth PNJ, as the transition width $d$ increases, the transmission probability of higher angular components diminishes. Notably, the transition width $d$ introduces a forbidden transport gap where transmission occurs exclusively through quantum tunneling\cite{allain2011klein}. 
\begin{figure}[!htbp]
	\includegraphics[width=\linewidth]{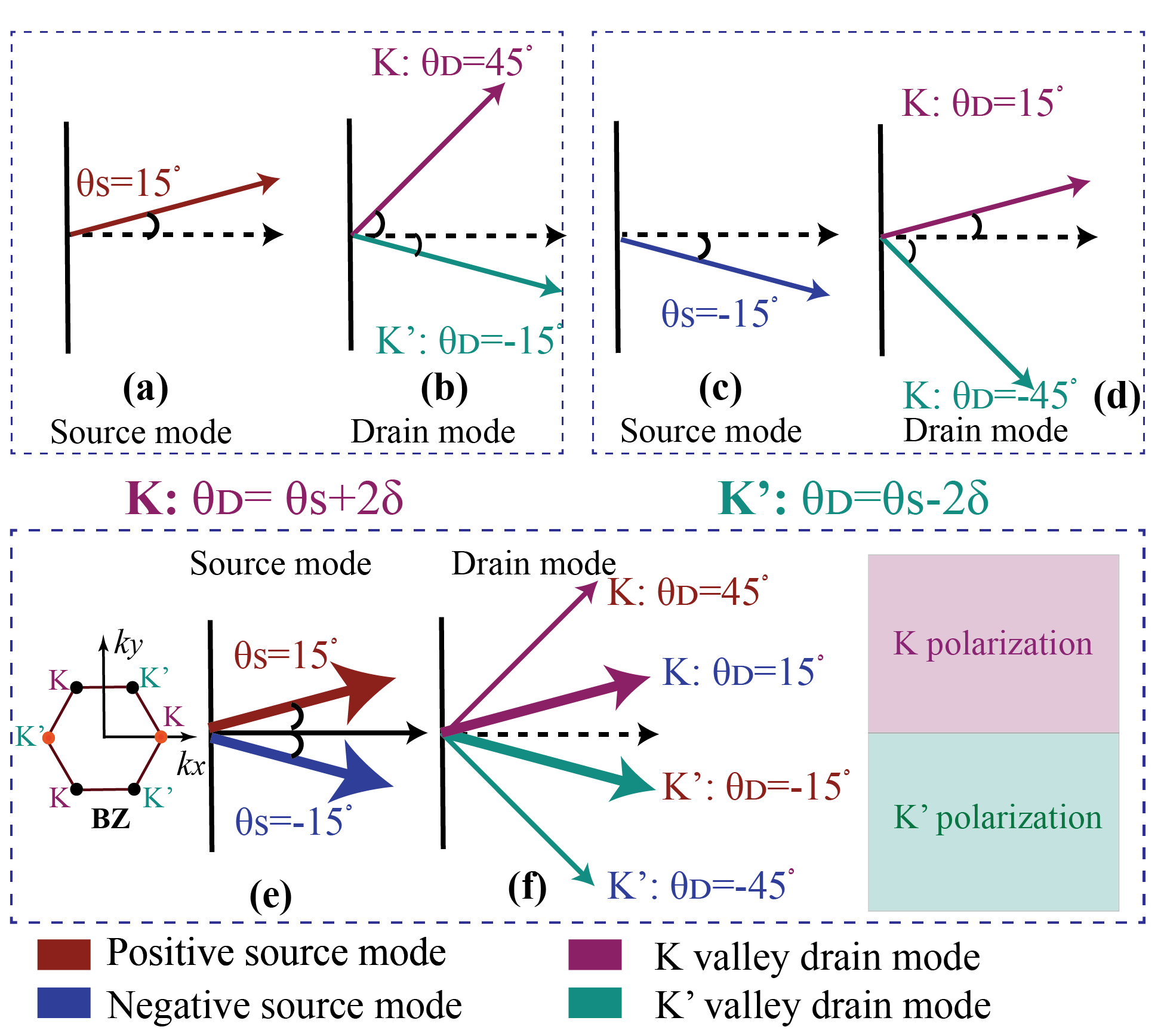}
    \caption{Working principle: Maximum transmission contributing modes in non-tilted BJ are normally incident mode  $\theta_s = 0$.  When the BJ is tilted with an angle $\delta=15^{\circ}$, the maximum contributing shifts to $\theta_S$ $\pm 15^{\circ}$ shown by maroon and blue colored arrows (a),(c), and (e). Superpositioned input modes with the equally probable $K$ and $K’$ components when incident on tilted junction gets scattered differently. The pseudo-spin conserved modes at the drain side for $K$ and $K’$ are different, given by the relation K/’=$ \theta_D$=$\theta_S$ $\pm \delta$}
    \label{fig:Fig_3}
\end{figure}\\
\indent In this study, we explore the functionality of tilted PNJs, which is built upon the foundation of normal PNJs, both exhibiting a smooth transition width. Considering normal PNJ, the incoming wave function from the input lead is a blend of $K$ and $K'$ wavevectors, each with an equal probability. When this wave function undergoes refraction across the junction, it maintains a combination of valley components on the transmitted side due to the uniformity of the potential profile along the $y$-axis, ensuring conservation of transverse wave components $k_y$. Consequently, both $K$ and $K'$ valley components persist equally, limiting distinct valley densities and distributions in real space.\\
\indent The potential profile in a normal PNJ aligns perpendicular to the lattice, maintaining translational invariance along the $y$-axis. This configuration ensures the conservation of the transverse wavevector component $k_y$ across the junction, facilitating particle transmission.\\
\indent When a normal PNJ is tilted at an angle $\delta$ relative to the $y$-axis, it forms a tilted PNJ, as depicted in Fig.\ref{fig:Fig_1} (a). In the tilted PNJ, the maximum contributing mode shifts from $\theta_s =0^{\circ}$ to $\theta_s = \pm \delta$. Consequently, the tilt in the junction shifts the angular transmission range towards a higher angular spectrum.\\
\indent In a tilted PNJ, the transverse wave component to the tilted junction equals $k_y$$\sin{\delta}$. The valley-degenerate source modes move in $\pm \theta_{s}$ directions. When encountering the tilted junction, they scatter into two possible pseudospin-conserved modes on the drain side, as depicted in Fig.\ref{fig:Fig_3}. The relationship between pseudo-spin conserved drain modes for $K$ (magenta-colored arrow) and $K'$ (green-colored arrow) components corresponding to the source mode $\theta_s$ and drain mode $\theta_D$ is given by\cite{PhysRevB.80.155406},
\begin{eqnarray}
K: \theta_D=\theta_s +2\delta, K': \theta_D=\theta_s-2\delta
\end{eqnarray}
\indent Considering the case when the tilt angle $\delta =15^{\circ}$, the maximum contributing source modes are now $\theta_s =\pm 15^{\circ}$ as shown in Fig.\ref{fig:Fig_3} (a) and (c). The $\theta_s =+15^{\circ}$  is shown by red-colored arrow (a) and the $\theta_s =-15^{\circ}$  by blue-colored arrow (c). The wavefunction superposed with the valley $K$ and $K'$ wave components transmits the $K$ and $K'$ components to distinct pseudospin-conserved drain modes when incident on a tilted junction. \\
\indent The pseudospin-conserved drain modes for an incident angle of $\theta_s=+15^{\circ}$ result in $\theta_D=45^{\circ}$ for the $K$ component and $\theta_D=-15^{\circ}$ for the $K'$ component. Similarly, at $\theta_s=-15^{\circ}$, the drain modes are $\theta_D=+15^{\circ}$ for $K$ and $\theta_D=-45^{\circ}$ for $K'$. In both cases, the maximum contributing modes remain at $\theta_d=\pm 15^{\circ}$. Since $K$ equals $-K'$, the valley components move in the same direction towards the drain. \\
\indent Fig.\ref{fig:Fig_3}(e) illustrates the most probable transmitting modes for a tilt angle of $\delta = 15^{\circ}$ (depicted by thicker arrows). In Fig. \ref{fig:Fig_3}(f), the corresponding pseudo-conserved $K$ and $K'$ modes for the given source modes are shown based on the relationship between $K$ and $K'$ for drain modes. Notably, it is observed that the modes most probable for transmission align closely with the angle of the tilt. Therefore, when $\theta_s = 15^{\circ}$, it contributes to $K'$ at $\theta_D = -15^{\circ}$ (depicted in thicker green), while $\theta_s = -15^{\circ}$ contributes to $K$ at $\theta_D = 15^{\circ}$ (depicted in thicker magenta). The thinner arrows in the graphical representation indicate that the components associated with the valley have a lower probability of transmission for a given source mode. Analyzing the component distribution reveals a prevalence of $K$ components in the upper half of the angular spectrum, contrasting with the dominance of $K'$ components in the lower half.\\
\indent In the cited work, \cite{PhysRevB.80.155406}, a parallel observation is noted. The authors delve into the impact of a titled NPJ on an edge-populated current. They attribute this effect to the negative refraction property of NPJ, without explicitly observing any valley splitting effects. The valley splitting effect observed in our work stems from the divergent configurations of the PNJ compared to the cited study. The PNJ in the cited work employs a flip configuration, inducing negative refraction or convergence toward the drain side. In contrast, our configuration fosters positive refraction or divergence, ensuring the separation of valley components.\\
\indent To make the most of this valley-specific angular spectrum, the designed device incorporates two output leads, each dedicated to one of the valleys. The subsequent section will delve into the outcomes across various tilt angles $\delta$ and transition widths.\\
\subsection{Varied tilt angles and transition widths: Impact on valley splitting}
\indent The tilted PNJ is defined by several parameters, encompassing the tilt angle $\delta$, transition width $d$, the longitudinal length of the junction $L$, and the separation distance between the tilted junction concerning the input and output leads (discussed in Appendix).\\
\indent In our simulations, we primarily focused on the tilt angle  $\delta$ and transition width $d$, which dictates an effective separation equal to $d/\cos\delta$. The longitudinal length $L$ of the junction varies with the tilt angle $\delta$ while maintaining a constant device width $W$, expressed as $L = W\cos{\delta}$. We meticulously maintained the distance between the lower end of the junction and the input lead $x_1$ within a minimal range, approximately between $10-15~nm$. Simultaneously, the distance $x_2$ between the upper end of the junction and the output lead, as illustrated in Fig. \ref{fig:Fig_1}(a), exhibited relatively stable values, ranging around $70-90~nm$ across different tilt angles. Consequently, an increase in the tilt angle resulted in a proportional increase in the overall device length.\\
\begin{figure}[!htbp]
    \includegraphics[width=\linewidth]{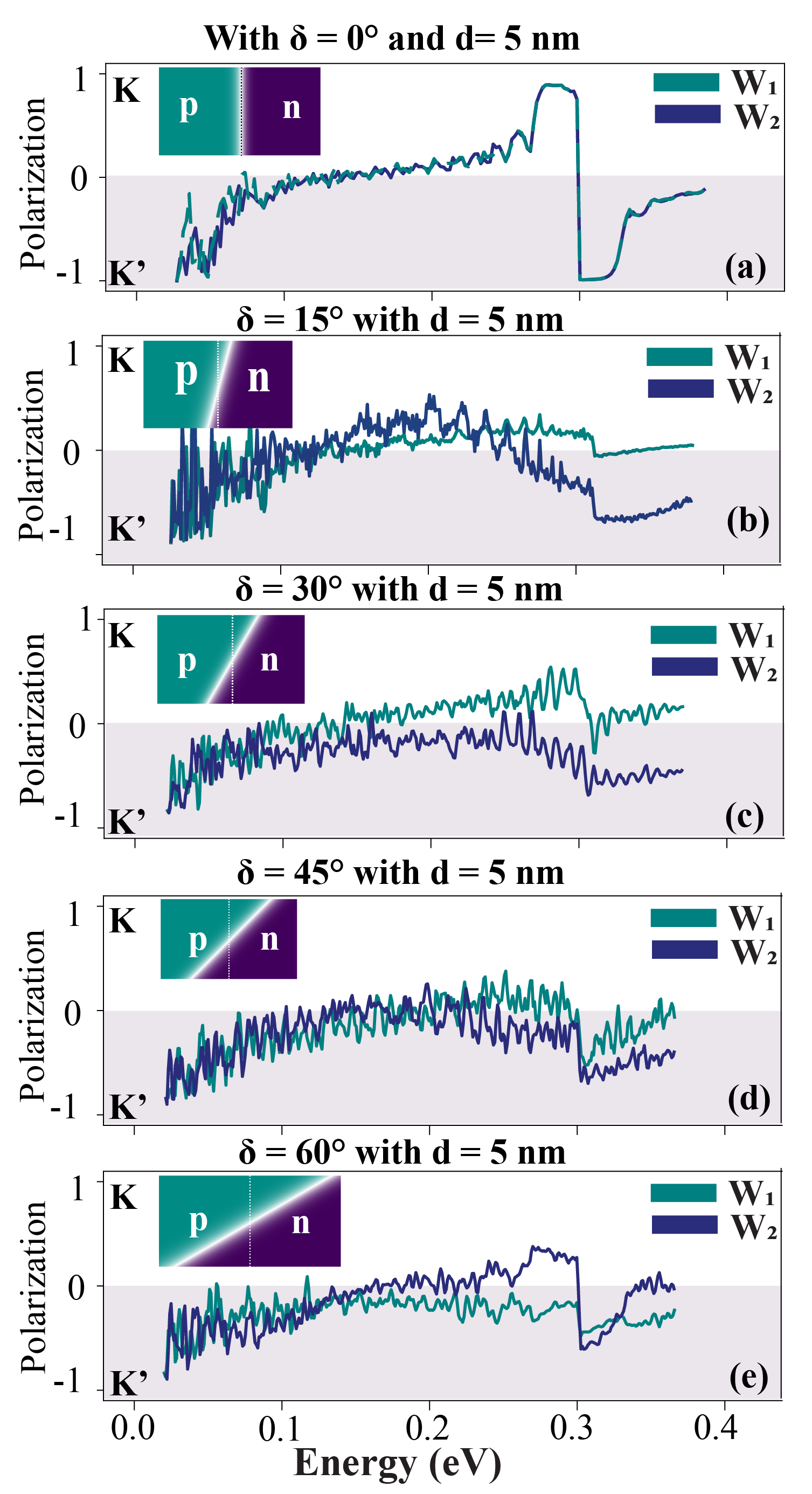}
    \caption{Polarization plot for different tilt angles with constant transition width $d=5nm$ across $n$ and $p$ region: (a) polarization for non-tilted/ straight junction over the energy range.  At the carrier energy $E =0.3~eV$, the peak in polarization is observed as an effect of the valley filter effect caused by the anti-zigzag configuration. The value of d is kept constant for all other tilt angles. (b) With a tilt angle of $\delta = 15^{\circ}$, the polarization curves experience a splitting after crossing the mid-barrier energy, reaching its peak at the barrier energy. Subsequently, there is a slight descent in polarization beyond this energy level. (c)  With $\delta= 30^{\circ} $, the polarization increases with the energy, and after crossing barrier energy it gets constant.  (d) For $\delta = 45^{\circ}$ no significant polarization is observed. Similarly, for $\delta =60^{\circ}$(e)the polarization curve approaches the non-tilted curve to some extent. Considerable valley splitting is observed for lower values of tilt angles.}
     \label{fig:Fig_4}
\end{figure}
\begin{figure*}[!t]
    \centering
    \includegraphics[width=0.95\textwidth]{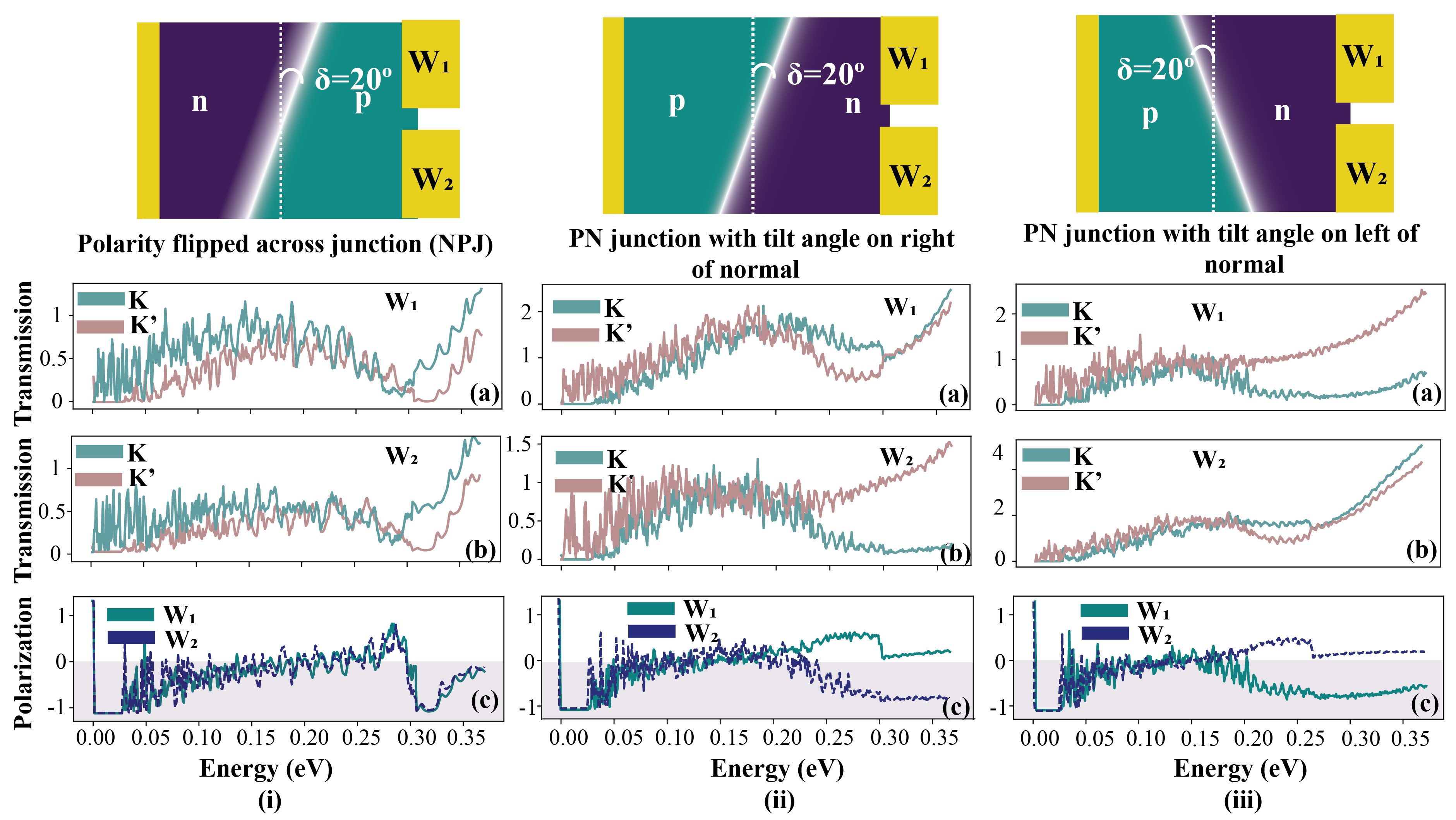}
    \caption{showcases transmission and polarization plots for three scenarios: (i) Flipped polarity across the junction, i.e., NP junction. (ii) The same conditions as discussed in the paper, featuring a tilted PNJ with the junction inclined to the left of the normal. (iii) A tilted PNJ with the junction inclined to the right of the normal. For all cases, the transition width remains constant at $10 \text{nm}$, and the tilt angle $\delta$ is consistent at $20^\circ$. (a) and (b) display the transmission of $K$ and $K'$ in $W_1$ and $W_2$, respectively. (c) illustrates the polarization curve in both output leads. In case (i), observations from transmission plots (i)(a) and (b) indicate no significant differences in valley components, a finding validated in (c) as it closely follows the curve of a normal PNJ junction without displaying any polarization. Moving to case (ii), where the PN junction is tilted to the left of the normal axis, observations from (ii)(a) and (b) reveal evident transmission differences in $K$ and $K'$ transmissions within $W_2$, while no polarization is observed in $W_1$ (c). In case (iii), where the junction is tilted to the right of the normal axis, observations from (iii)(a) and (b) show nearly the reverse of case (ii). Here, evident valley contrast is observed in $W_1$, while negligible polarization is noticed in $W_2$ }
     \label{fig:Fig_5}
\end{figure*}
\indent Initially, the investigation entailed altering the tilt angle while maintaining a constant transition width $d=5~nm$. The different values of tilt angles have been considered in the range of $0^{\circ}$ to $60^{\circ}$ relative to the $y$-axis to explore the polarization variations of valley components $K$ and $K'$ across two output leads. The incoming fermion energy spectrum spans from $0.0~eV$ to $0.36~eV$ across all the tilt angles. The transmission plots for each valley component across the output leads are provided in the Appendix part, covering the complete tilt angle and energy range.\\
\begin{figure*}[!t]
    \centering
    \includegraphics[width=0.95\textwidth]{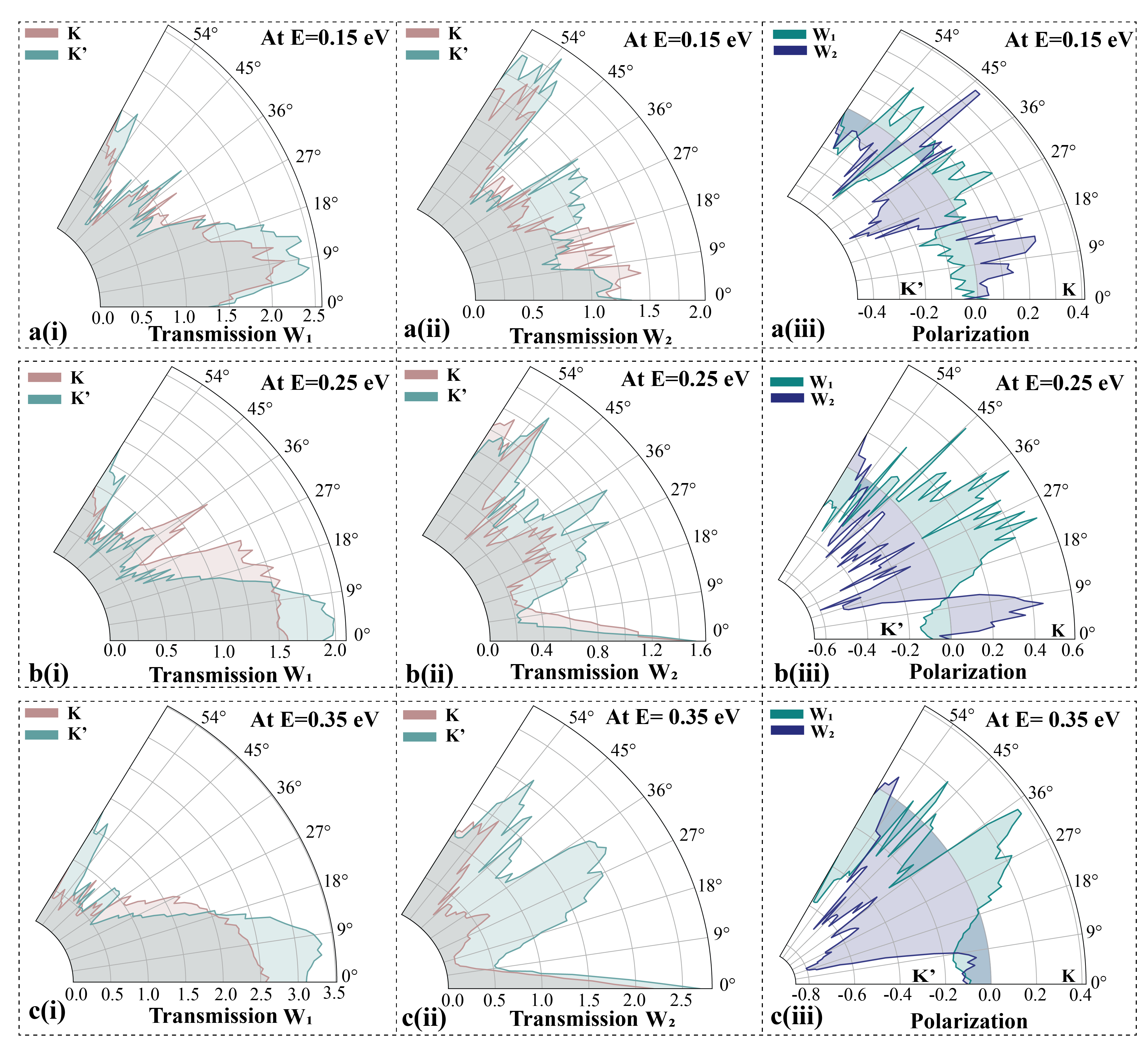}
    \caption{Transmission and polarization trend over the tilt angle spectrum at different energies: The plots $\bold{(i)}$ and $\bold{(ii)}$ are the $K$ and $K'$ transmission plots for the output leads $W_1$ and $W_2$ respectively, and $\bold{(iii)}$ is the polarization plot at an energy $E$ for tilt angle spectrum ranging from $0^{\circ}$ to $60^{\circ}$.
    $\bold{Case (a)}$ corresponds to a carrier energy $E$ of $0.15 ~eV$. In a(i), transmission is higher at lower tilt angles and gradually degrades approaching the mid-tilt angle spectrum. No significant valley contrast is observed at this energy. In a(ii), high transmission is observed at higher tilt angles, with a slight valley contrast that flips polarity at certain intervals across the tilt angle range. This flipping pattern is evident in a(iii), where the polarization magnitude is small and changes simultaneously across the output leads at intervals.$\bold{Case (b)}$ corresponds to a carrier energy $E$ of $0.25~eV$, where the device functions as a bipolar junction. For $W_1$ in b(i), the trend is similar to a(i), with higher transmission and low valley contrast in the lower tilt angle range. In b(ii), the transmission value peaks near the normal angle and higher valley contrast is observed in the middle tilt angle range. The polarization plot in b(iii) illustrates this scenario clearly, where the middle tilt angle range exhibits substantial transmission and high valley contrast.$\bold{Case (c)}$ corresponds to a unipolar junction when the carrier energy is $E=0.35$ eV, exceeding $V_0 =0.3$ eV. The effects observed in Case (b) become more profound, with high transmission and low valley contrast in c(i), and high valley contrast with considerable transmission observed in the mid-tilt angle range in c(ii). The polarization plot illustrates these effects in (i) and (ii), indicating that the optimal condition for operation is in the mid-tilt angle range in c(iii).}
     \label{fig:Fig_6}
\end{figure*}
\indent For a better understanding, we commence with the examination of valley polarization behavior in a normal PNJ, serving as a reference point. The valley polarization curve for the normal PNJ for a given device setup is depicted in \ref{fig:Fig_4}(a). The green and blue curves correspond to the polarizations in output leads $W_1$ and $W_2$, respectively.  Positive values indicate polarization in $K$, while negative values signify polarization in $K'$. Notably, the first observation is that the polarization is entirely identical across the two output leads. Furthermore, distinct peaks in the polarization curve are observed on either side of the barrier energy ($V_0$=0.3~eV). These peaks are attributed to the valley filter effect resulting from the anti-zigzag configuration\cite{akhmerov2008theory,cresti2008valley,PhysRevB.107.205415} employed in the simulation.\\
\begin{figure*}[!t]
    \centering
    \includegraphics[width=0.95\textwidth]{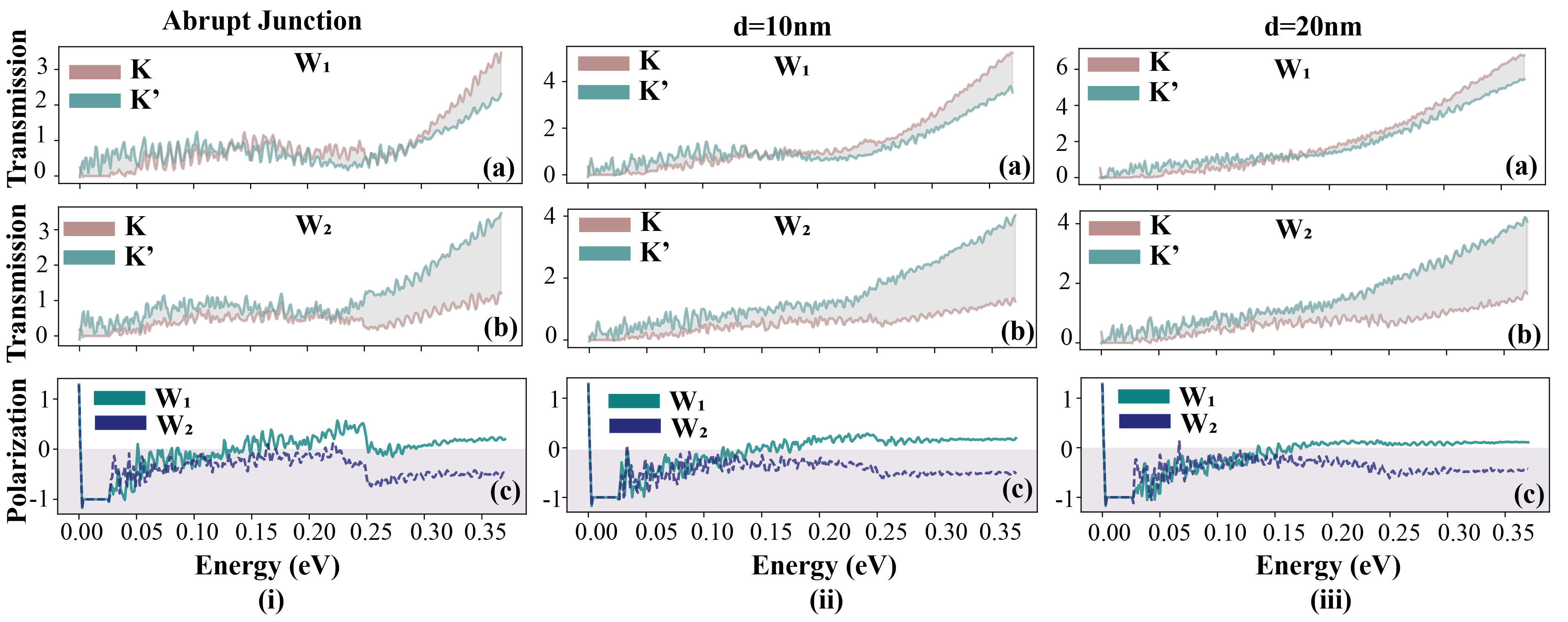}
    \caption{Variations in Transmission and Polarization with Transition Width ($\delta =30^{\circ}$): For a tilted junction with varying transition widths ($d$), transmission values for $K$ and $K'$ in $W_1$ and $W_2$ are depicted in plots (a and b) respectively. Plot (c) showcases the polarization curves in $W_1$ and $W_2$. In the scenario where the tilted junction is abrupt ($d=0nm$), below the barrier energy of 0.25 eV, transmission values are minimal. As the fermion energy approaches the barrier energy, polarization across the output lead begins to rise, indicating a valley filter effect at the barrier energy. Post this phase, the polarization stabilizes. With $d=10nm$ (case ii), compared to case (i), two notable differences are observed: a higher maximum transmission value $\bold{ii (a and b)}$ and a more flattened polarization curve after the barrier energy(c). Increasing the transition width to $d=20nm$ (case iii) further emphasizes the previously observed effects. Here, the transmission rates are higher overall, and the polarization curves exhibit greater flattening post-barrier energy }
     \label{fig:Fig_7}
\end{figure*}
\indent The plot presented in Fig.\ref{fig:Fig_4}(b, c, d, and e) depicts polarization curves corresponding to four tilt angles: $\delta = 15^{\circ}$, $30^{\circ}$, $45^{\circ}$, and $60^{\circ}$ with respect to the $y$-axis, while maintaining a constant $d$ value of $5~nm$. The polarization dynamics exhibit significant variations at a $\delta = 15^{\circ}$ tilt angle along the $y$-axis, as illustrated in Fig. \ref{fig:Fig_4}(b). As the energy increases, the polarization curves begin to diverge beyond the mid-potential barrier energy ($V_0/2$), manifesting opposite polarities. These values peak after crossing the barrier energy at $0.3~eV$ and experience a slight decrease thereafter.\\
\indent For $\delta = 30^{\circ}$ Fig.\ref{fig:Fig_4}(c), a distinction from the previous scenario is observed as the polarization curve starts to separate at lower energies, maintaining a constant trend after surpassing the barrier energy.
At $\delta = 45^{\circ}$, both curves exhibit erratic behavior, oscillating around zero polarization values. After the barrier energy, a slight polarization or curve splitting is observed.     
For $\delta = 60^{\circ}$, no significant polarization is evident for either output lead. However, at the barrier energy, the characteristic polarization, attributable to the filter effect, appears at varying levels in each case across the output leads.\\
\indent In the previous section, we emphasized the significance of the sequencing of doped regions when forming the junction. Here, we present the transmission pattern and polarization for the flipped polarity of doped regions across the junction. Additionally, we illustrate the reversal of the tilt angle direction with respect to the $y-axis$ compared to the direction considered in this study. To maintain a common reference, we keep the tilt angle value constant at $\delta = 20^{\circ}$ and $d$ = 10nm. Schematics are provided for each configuration, with plot (a) and (b) showing the transmission curve for $K$ and $K’$ in $W_1$ and $W_2$, respectively, while plot (c) offers insight into the polarization in two output leads.\\
\indent In the scenario where doping polarities across the junctions are flipped, as depicted in Fig. \ref{fig:Fig_5} case (i), with an n-doped region followed by a p-doped region, it is evident from (i)(a) and (b) that both valleys exhibit nearly identical transmission patterns. This consistency is also apparent in (c) with the polarization curve. The polarization associated with the valley filter effect appears near the barrier energy, yet the splitting of valleys across output leads is absent. This consistent outcome holds for any tilt angle, although not explicitly shown here. The absence of valley polarization occurs because, on the left side of the junction, which also serves as a refracted zone, the presence of $p-type$ doping induces a converging effect. Consequently, this effect tends to focus the separated valley fermions, causing them to merge.\\
\indent In Fig.\ref{fig:Fig_5} case (ii), the sequence of doped regions across the junction is the same as predominantly considered in the paper, with the inclination of the junction on the right side of the $y$-axis. The transmission characteristics in (a) and (b) are consistent with the findings outlined in the previous section. Notably, $W_2$ exhibits a predominance of $K'$ with no substantial variance observed in $W_1$. The polarization plot (c) confirms the polarization observed in $W_2$ and maintains a consistent level of polarization beyond the barrier energy.\\
\indent In Fig.\ref{fig:Fig_5} case (iii), where the tilt direction shifts to the left side of the normal axis, the transmission pattern observed in case (ii) across $W_1$ and $W_2$ flips in the output leads, as evident in case (iii)(a) and (b). Notably, in case (iii), there is a significant polarization observed in $W_1$ compared to $W_2$.\\
\indent A notable similarity exists between Fig.\ref{fig:Fig_5} case (ii)(a) and case (iii)(b) where no polarization is detected, and the transmission of both $K$ and $K'$ components is comparable. Moreover, the transmission values in these scenarios are higher than their counterparts, primarily due to the tunneling effect.\\
\indent For a detailed examination of the relationship between energy and valley polarization, Fig.\ref{fig:Fig_6} presents polar plots showcasing valley polarization in two output leads across the entire range of considered tilt angles at three specific energies. These energies—$E=0.15~eV$ and $E=0.25~ eV$—fall before the barrier energy, while $E=0.35 ~eV$ is positioned after it.\\
\indent In Fig.\ref{fig:Fig_6}, $\bold{(i)}$ showcases the transmission of $K$ and $K'$ in $W_1$, while $\bold{(ii)}$ illustrates the transmission in $W_2$. $\bold{(iii)}$ represents valley polarization in both $W_1$ and $W_2$.To facilitate explanation, we categorize the angular spectrum as follows: the lower range from $0^\circ$ to $20^\circ$, the middle range from $20^\circ$ to $40^\circ$, and the higher range from $40^\circ$ to $60^\circ$.\\
\indent Considering $\bold{case\ a}$: At an energy of $E=0.15~eV$, the transmission in Fig.\ref{fig:Fig_6} $  \bold{a(i)}$  exhibits higher values at lower tilt angles, with nearly equivalent presence of both valley components over an angular range. In Fig.\ref{fig:Fig_6}$\bold{a(ii)}$, higher tilt angles show higher transmission values with almost the same equivalence of valley components. Within the lower angular range, there is a slight dominance of a valley component that flips around approximately $20^{\circ}$ tilt. These observations are confirmed in the polarization plot Fig.\ref{fig:Fig_6}$\bold {a(iii)}$, where a small polarization is evident at this energy. The polarity across the output leads flips at approximately $20^{\circ}$ and $40^{\circ}$ tilt.\\
\indent Moving to Fig.\ref{fig:Fig_6} $\bold{case\ b}$: At $E=0.25~\text{eV}$, as discussed earlier, Fig.\ref{fig:Fig_6} 
$\bold{b(i)}$ at a lower tilt angle range, the transmission is higher, like the previous, but a slight dominance of $K'$ is observed. Fig.\ref{fig:Fig_6} $\bold {b(ii)}$ a high transmission is observed for the tilt angles nearby normal. There is a decrease in transmission then after. In the mid-range of angles, the transmission increases with the dominance of $K'$ which flips the polarity at a higher angle. A clear understanding can be achieved from the polarization plot, Fig.\ref{fig:Fig_6} $\bold{b(iii)}$, the middle tilt angle range shows maximum polarization with opposite polarities in each output lead.\\
\indent Lastly, Fig.\ref{fig:Fig_6} $\bold{case\ c}$: At $E=0.35~\text{eV}$, within the unipolar regime, the trend observed in the previous case, Fig.\ref{fig:Fig_6} $\bold{case\ b}$, becomes more evident and enhanced here. Higher transmission at lower tilt angles in Fig.\ref{fig:Fig_6} $\bold{c(i)}$ and higher transmission in Fig.\ref{fig:Fig_6} $\bold{c(ii)}$ post the maximum transmission in Fig.\ref{fig:Fig_6}$ \bold{c(i)}$ goes down. From polarization plot Fig.\ref{fig:Fig_6} $\bold{c(i)}$, the maximum polarization is observed in mid-spectrum for a given range with opposite polarities across the output leads.\\
\indent Based on the preceding discussion, the optimized conditions for maximizing valley polarization can be summarized as follows: 1. Tilt angles should be within approximately between $20^{\circ}$ to $40^{\circ}$. 2. The degree of polarization increases with rising energy, peaking at the barrier energy, and maintains a constant value thereafter.\\
\indent Now, we elucidate the plausible explanations for the observed effects in the results. The absence of substantial polarization in the upper half tilt angle spectrum can potentially be explained by mode mixing and interference effects. With an increase in the tilt angle, both the longitudinal length and area of the junction expand. The smooth transition width leads to a transmission gap, promoting transport via quantum tunneling. Consequently, mode mixing and interference occur in the interface region, effectively nullifying the pseudospin conserved transport\cite{PhysRevB.80.155406,yang2012conductance}.\\
%\indent The constancy in polarization beyond the barrier energy, despite the transition to unipolar junctions, emerges due to the asymmetric energy levels prevailing across the junction. The energy of fermion before is ($V_0$-$E$) and after junction being $n-type$ it is at positive $0.3~eV$ above the incoming fermion energy.  As the transmitting modes encounter a higher positive refractive index on one side, they tend to generate divergent beams on the opposite side of the junction. Additionally, the existence of a critical angle leads to total internal reflection, resulting in fewer modes transmitting through the junction, which then experience significant diffraction. This limits the further enhancement of the valley polarization effect and the influence remains consistent across both valley components.\\
\indent We explored the impact of transition width on transmission values and polarization in a device by examining three variable values: $d=0~nm, ~10~nm, ~20~nm$ while maintaining a single tilt angle of $\delta=30^{\circ}$. In Fig. \ref{fig:Fig_7} $\textbf{(a and b)}$, the transmission of $K$ and $K’$ components in output leads $W_1$ and $W_2$ is depicted, while $\textbf{(c)}$ details the polarization in these output leads.\\
\indent $\textbf{Case\ i:}$ With $d=0~nm$, representing an abrupt junction, the transmission values before $V_0=0.25~eV$ are lower due to the expected potential barrier Fig.\ref{fig:Fig_7}$\textbf{(i and ii)}$. Valley contrast is not observable initially. However, as the carrier energy approaches the barrier height, slight peaks in polarization are observed in Fig.\ref{fig:Fig_7}$\textbf{i(c)}$ due to the valley filter effect \cite{akhmerov2008theory}. Beyond this energy, significant polarization is noted in $W_2$ and a smaller effect in $W_1$.\\
\indent $\textbf{Case ii:}$ At $d=10~nm$, the higher transmission values are evident in the transmission curves Fig.\ref{fig:Fig_7} $\textbf{ii (a and b)}$, displaying a noticeable difference in $K$ and $K’$ transmissions at lower energies $\textbf{ii(b)}$. Surprisingly, the maximum transmission values are higher than the previous case, countering expectations, and the valley filter effect diminishes. This is explained later. Moreover, the transmission in $W_1$ tends toward an equivalence between the transmission values of $K$ and $K'$.\\
\indent $\textbf{Case iii:}$ Increasing $d=20~nm$ accentuates the effects observed in $\bold{case(ii)}$. Maximum transmission values are even higher in Fig.\ref{fig:Fig_7} $\textbf{iii(a and b)}$ than in the previous cases. The valley filter effect at $V_0$ in $\bold{iii(c)}$ diminishes further. The valley component separation in $W_2$ has noticeably increased, while in $W_1$, it remains negligible. The same can can be observed in polarization Fig.\ref{fig:Fig_7} $\textbf{iii(c)}$\\
\indent The observed rise in transmission value with increasing $d$ might seem counter-intuitive. Normally, in a non-tilted PNJ when $d$ increases, it signifies an expansion in the length scale over which potential gradually varies at an atomic level. The increase in $d$ leads to an induced transport energy gap, favoring transmission solely through quantum tunneling. Crucially, this increase in $d$ also corresponds to the suppression of higher angular modes, thereby reducing the transported angular bandwidth.\\
\indent When the PNJ junction is tilted, the longitudinal length of the junction, denoted as $L = d * \cos{\delta}$, increases, expanding the physical junction area. Intuitively, a broader transition area typically corresponds to a further decrease in the transmission value. However, the current observation contradicts this expectation, showing an increase in transmission values despite the angular lobes shifting to higher angular spectra due to the tilting of a junction where suppression of modes is evident.\\
\indent The current findings correlate with the initial observation in tilted PNJ setups, showcasing an increase in conductance with the rise in tilt angle up to the threshold angle $\delta_{th}$, followed by a decline \cite{sajjad2012manifestation, PhysRevB.80.155406}. Interestingly, increasing the transition width $d$ can postpone this threshold angle occurrence. Consequently, increasing $d$ shifts the threshold angle to higher angular values, enabling higher transmission probabilities. The study also advocates the utilization of tilted PNJ configurations to enhance conductance.\\
\indent The observed rise in conductance in tilted junctions is attributed to specular edge scattering events. This increase arises due to enhanced forward scattering events \cite{sajjad2012manifestation}. Specifically, electrons above the critical angle striking the edge reflects back to the junction and then transmit through it, adhering to Snell’s Law. This phenomenon ensures the preservation of chiral tunneling and valley trajectory. The edge scattering events are prominent with the long channel devices, as evident in this work.\\
\subsection{Effects of an edge disorders:}
\begin{figure}[!htbp]
    \includegraphics[width=\linewidth]{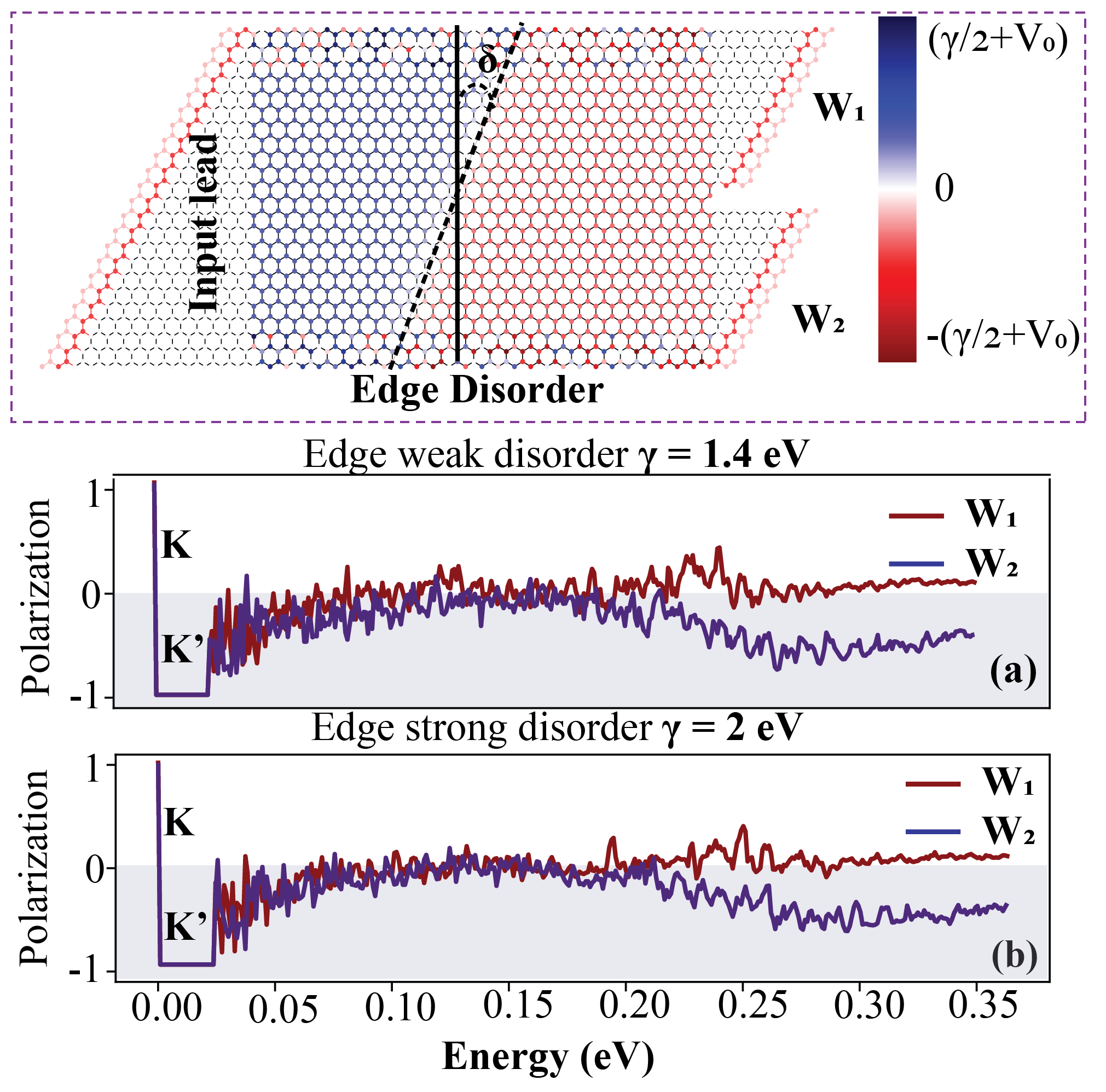}
    \caption{Valley polarization plots were generated by introducing edge Anderson short-range disorder of different magnitudes:(a) Weak edge disorder was represented by a strength of $\gamma = 1.4eV$. (b) Strong edge disorder was simulated at a strength level of $\gamma = 2 eV$. Notably, the polarization persisted despite the presence of both weak and strong disorder strengths at the edge.}
     \label{fig:Fig_8}
\end{figure}

\indent In the previous section, as stated, the enhancement in pseudospin-conserved transmission in tilted PNJ results from speculative edge scattering. In practical situations, disorders at the edges are frequently observed.\cite{mucciolo2010disorder}. To assess the performance of the proposed device model under such conditions, we introduced Anderson short-range edge-type disorder with varying strengths.\\
\indent A short-range Anderson disorder along the device edges, characterized by normal distribution-based potential fluctuations is introduced to the onsite energies \cite{xiong2007anderson, schubert2009anderson}. This disorder is termed short-range due to variations occurring on a scale smaller than the lattice constant $a$. The resulting short-range potential fluctuations create local PNJs. In graphene with zigzag edges, these localized PNJs induce inter-valley scattering and zero backscattering, locally breaking time-reversal symmetry (TRS) while maintaining its overall preservation \cite{lherbier2008transport, wakabayashi2007perfectly, wurm2012symmetries}.\\
\indent We incorporated the Anderson short-range disorder uniformly along the edges, leading to uncorrelated disorder potentials. The final onsite potential is defined as:
\begin{eqnarray}              
 \epsilon_{e}= \epsilon_i + \delta\epsilon_{e}.
\end{eqnarray}
Here, the subscript $e$ represents the edge disorder. The term $\delta\epsilon_{e}$ ranges within $[-\gamma/2, \gamma/2]$, where $\gamma$ denotes the strength of the disorder potential, and $\epsilon_i$ signifies the onsite energy due to electrostatic doping. For edge disorder, we considered three atomic row thicknesses on both edge sides and introduced the scattering potential over $p$ and $n$-electrostatically doped regions as shown in Fig.\ref{fig:Fig_8} lattice schematic.\\
\indent In the context of a hopping integral value of $t = -2.7$ eV, disorder strengths within the range of $[0.26-1]$ eV are regarded as relatively weak, while those ranging between $[1.3, 2.7]$ eV are considered strong. Our focused investigation targeted two specific disorder strengths: $\gamma = 1.4$ eV, falling into the weak category, and $\gamma = 2$ eV, classified as high. To accommodate the diverse spatial distributions inherent in the disorder potential, we conducted simulations encompassing thirty distinct configurations for each strength level. For simulation purposes, we considered a tilt angle of $\delta=30^\circ$ and a transition width of $d=5~nm$.\\
\indent Fig.\ref{fig:Fig_8} (a and b) illustrates the polarization plots concerning edge disorder characterized by weak and strong strengths. The observation indicates a slight degradation in polarization with an increase in disorder strength. However, notably, the polarization predominantly withstands the influence of the edge disorder \cite{areshkin2007ballistic}.\\
\indent This resilience to edge disorder can be attributed to the fact that bulk modes are less susceptible to such disorders, while edges exhibit a large local density of states (LDOS) within a narrow energy spectrum close to the charge neutrality point due to their zigzag edge configuration \cite{areshkin2007ballistic}. Thus, they tend to circumvent the edges, thereby rendering these regions impervious to the effects of edge disorder\cite{fujita1996peculiar,nakada1996edge}.

\section{Conclusion} \label{Sec4}
\indent In this work, we introduced a multi-terminal monolayer graphene device featuring a tailored tilted P-N junction (PNJ) designed specifically to induce valley splitting within real space. Typically, graphene exhibits isotropic transport across the continuum energy spectrum for both energy-degenerate valleys. However, achieving valley-specific transport becomes imperative to delineate discernible valley contrasts. To achieve this, we integrated a tilted PNJ, thereby modifying the pseudospin-conserved modes upon refraction through the potential interface. This modification facilitated valley-specific tunneling, consequently establishing distinct valley-resolved transport patterns within the system. Our analysis delved into optimizing the doping profile and fermion energy to attain optimal operating conditions. The fermion energy at the junction dictated the junction type—be it unipolar or bipolar—thereby influencing the refractive properties of the system through Dirac-fermion optics. For device optimization, we scrutinized two critical parameters: the tilt angle $\delta$ and transition width $d$ across the junction. Notably, an increased transition width was found to enhance transmission across the device, attributed to specular edge scattering. Remarkably, our comprehensive analysis revealed the system remains resilient against the influence of Anderson short-range edge disorder. Significantly, this mechanism draws parallels with the anisotropic chiral tunneling observed in the Weyl-Dirac system, stemming from its tilted Dirac energy spectrum. The incorporation of a tilted PNJ into the isotropic Dirac system holds promise for inducing analogous anisotropic behaviors, potentially transforming isotropic Dirac systems to resemble tilted Dirac-Weyl semimetals. 
\begin{acknowledgments}
The authors acknowledge the Science and Engineering Research Board (SERB), Government of India, for Grant No. CRG/2021/003102, and the Ministry of Education (MoE), Government of India, Grant No.STARS/APR2019/NS/226/FS under the STARS scheme
\end{acknowledgments}

\appendix
\nocite{*}
\bibliography{apssamp}% Produces the bibliography via BibTeX.
\end{document}